\documentclass[conference]{IEEEtran}
\IEEEoverridecommandlockouts
\usepackage{cite}
\usepackage{amsmath,amssymb,amsfonts}
\usepackage{algorithmic}
\usepackage{graphicx}
\usepackage{textcomp}
\usepackage{xcolor}

\usepackage{booktabs}
\usepackage{hyperref}

\newcommand{\scipy}{\texttt{\texttt{SciPy}}~}

\newcommand{\sLLGS}{\mbox{s-LLGS}\ }

\begin{document}

\title{A Fokker-Planck Solver to Model MTJ Stochasticity}

\author{
\IEEEauthorblockN{Fernando García-Redondo}
\IEEEauthorblockA{\textit{Arm Ltd, Cambridge, UK} \\
fernando.garciaredondo@arm.com}
\and
\IEEEauthorblockN{Pranay Prabhat}
\IEEEauthorblockA{\textit{Arm Ltd, Cambridge, UK} \\
pranay.prabhat@arm.com}
\and
\IEEEauthorblockN{Mudit Bhargava}
\IEEEauthorblockA{\textit{Arm Inc, Austin, USA} \\
mudit.bhargava@arm.com}
}

\maketitle

\begin{abstract}
	Magnetic Tunnel Junctions (MTJs) constitute the novel memory element in STT-MRAM, which is ramping to production at major foundries as an eFlash replacement. MTJ switching exhibits a stochastic behaviour due to thermal fluctuations, which is modelled by  \sLLGS and Fokker-Planck (FP) equations. This work implements and benchmarks Finite Volume Method (FVM) and analytical solvers for the FP equation. To deploy an MTJ model for circuit design, it must be calibrated against silicon data. To address this challenge, this work presents a regression scheme to fit MTJ parameters to a given set of measured current, switching time and error rate data points, yielding a silicon-calibrated model suitable for MRAM macro transient simulation.
	
\end{abstract}

\begin{IEEEkeywords}
	STT-MRAM, MTJ, s-LLGS, Fokker-Planck
\end{IEEEkeywords}

\section{Introduction}

Magnetoresistive Random Access Memory (MRAM) devices have been actively explored during the last decade
\cite{Lee2018, Torunbalci2018a, Zhang2020, Boujamaa2020} as one of the most promising Non-Volatile Memory (NVM) technologies, with applications from embedded Flash replacement to future power-efficient caches in HPC systems. In particular, Spin-Transfer Torque MRAM (STT-MRAM) is being actively developed by foundries and integrated into 28nm generation CMOS Process Design Kits (PDKs) \cite{Boujamaa2020}. MTJ structures are complex multi-layered devices with their magnetization behavior described by the Landau-Lifshitz-Gilbert-Slonczewsky (LLGS) system~\cite{Donahue1999a, Ament2016, Garcia-Redondo2021}. Apart from the material parameters, external and induced magnetic fields, the evolution of MTJ magnetization $\boldsymbol{m}$ also has a stochastic nature, resulting from a field induced by thermal fluctuations.

This stochastic behavior, described by the \sLLGS equations with the addition of a thermal field component $\boldsymbol{H_{th}}$, influences the read and write operations in MTJ-based memories and induces non-deterministic bit errors which are traditionally expressed as Write/Read Error Rates (WER/RER). Even without considering parameter variation, the computation of WER/RER with \sLLGS simulations requires a large number of random walks, especially for the low error rates ($<<1ppm$) required for volume production. 

To alleviate this issue,  Stochastic Differential Equation (SDE) tools such as the Fokker–Planck Equation (FPE) statistically analyze the MTJ magnetization and provide a simplified solution with sufficient accuracy to analyze such error rates \cite{Butler2012, Xie2017, Torunbalci2018a}. Compared against a set of \sLLGS random-walk transient simulations, the FPE accurately evolves an initial MTJ magnetization probability through time based on the current and external fields. Instead of independent transients, the FPE computes the probability distribution of the magnetization at a given instant, thus capturing the statistical behavior of the MTJ cell.

This paper presents a framework for the characterization and calibration of MTJ stochastic effects. Section~\ref{sec:motivation} introduces the \sLLGS and FP equations and the proposed overall model framework. The numerical solution of the FPE has trade-offs between the computational load and accuracy of the solvers \cite{Xie2017, Daniel2021}. An FVM-based solver and an analytical solver are implemented and benchmarked in Section~\ref{sec:solvers}.

Circuit design with MTJ compact models requires accurate fitting to measured silicon data. Silicon measurement needs low error rates to be captured accurately from finite memory arrays without compromising test throughput in volume production, requiring high currents to allow extrapolation from higher, more easily measurable error rates. As a result, foundry data could consist of a set of data points with the error rate spanning orders of magnitude. Section~\ref{sec:case_study} presents a regression framework for fitting the complex set of MTJ parameters onto such heterogeneous data points through a case study with published foundry data. The fitted set of parameters is further calibrated with a set of thermal fitting coefficients feeding into a transient model~\cite{Garcia-Redondo2021}, showing how circuit designers can emulate the switching behaviour of stochastic corner MTJs corresponding to target WER values (say $0.5$, $10^{-6}$ and $10^{-8}$) for accurate power-performance characterization of a memory macro. Finally, Section~\ref{sec:conclusions} draws the conclusions from this work.

\section{s-LLGS and FPE Analyses}
\label{sec:motivation}
\begin{figure}[!t]
\centering
\includegraphics[width=0.9\columnwidth]{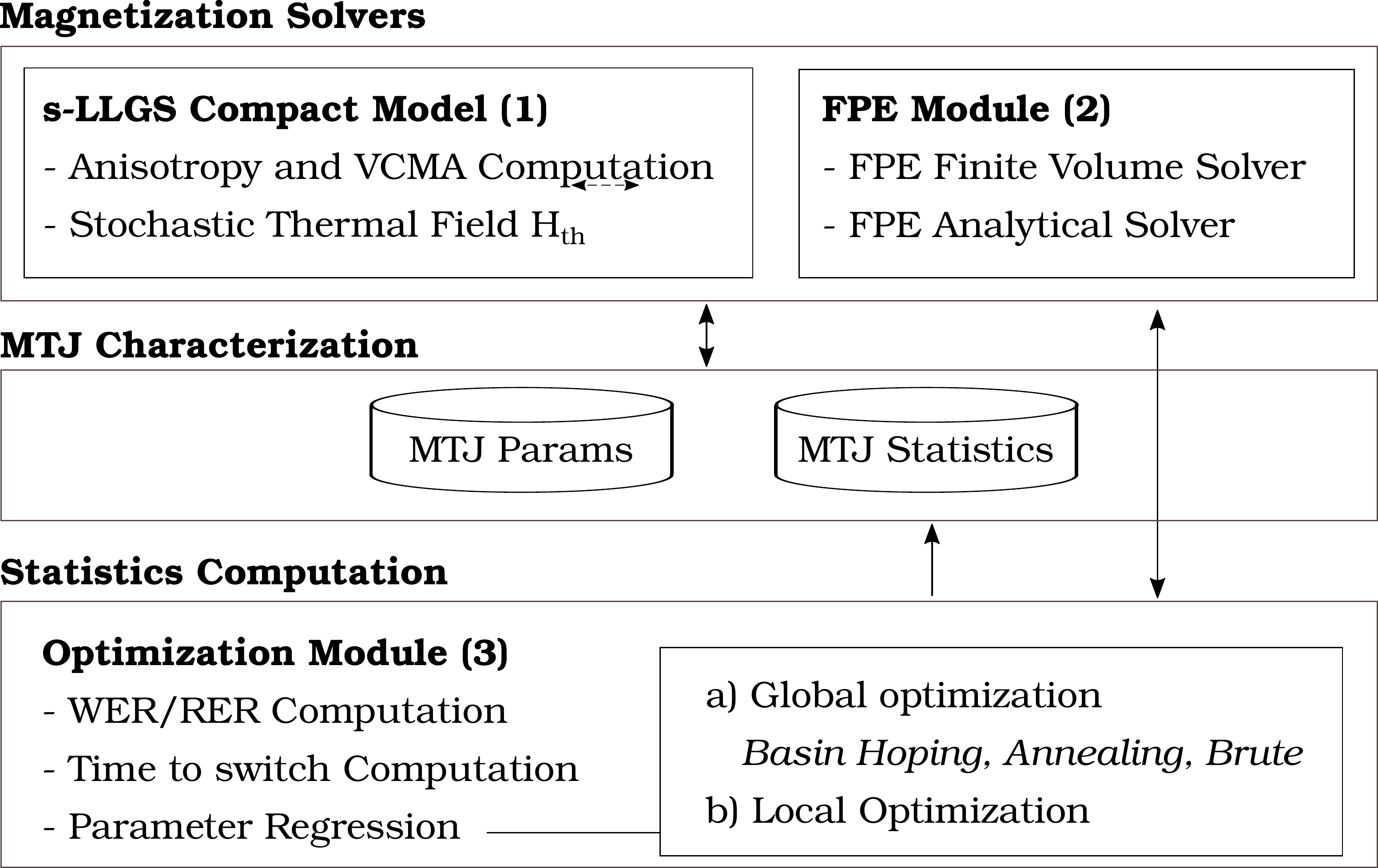}
\caption{
	Overview of the framework performing the MTJ characterization and statistical analyses: s-LLGS, FPE solvers and optimization modules.
}
\label{fig:proposal}
\end{figure}

\sLLGS and FPE-based analyses has been widely used to characterize MTJ switching. This section introduces the equations and presents the proposed FPE solvers. Figure \ref{fig:proposal} describes the proposed toolbox, consisting of an \sLLGS compact model \cite{Garcia-Redondo2021}, a FVM FPE solver, an analytical FPE solver based on~\cite{Tzoufras2018} and optimization tools using global minimization \cite{Virtanen2020}. The \sLLGS compact model computes stochastic random transient walks based on a set of MTJ parameters. FPE solvers evolve a known initial magnetization probability distribution $\rho_0(\theta)$, computing the magnetization probability  at a given time $\rho(\tau, \theta)$. Finally, the optimization module solves for MTJ statistics and fits parameters to achieve a target error rate.

\subsection{MTJ Stochasticity: Origin in s-LLGS}
The temporal evolution of MTJ magnetization $\boldsymbol{m}$ can be described as a monodomain nanomagnet influenced by external and anisotropy fields, thermal noise and STT \cite{Donahue1999a, Ament2016}, ruled by the \sLLGS equations \cite{Donahue1999a} in I.S.U.
\begin{eqnarray}
	\frac{d \boldsymbol{m}}{dt} = & -\gamma' \boldsymbol m \times \boldsymbol{H_{eff}} \nonumber + \alpha \gamma' \boldsymbol{m} \times \frac{d \boldsymbol{m}}{dt} \label{eq:sllgs} \\
		& + \gamma' \beta \epsilon ( \boldsymbol{m} \times \boldsymbol{m_p} \times \boldsymbol{m}) - \gamma' \beta \epsilon' (\boldsymbol{m} \times \boldsymbol{m_p})
\end{eqnarray}
where $\alpha$ and $\gamma$ are the Gilbert damping factor and gyromagnetic ratio respectively, related by $\gamma' = \frac{\gamma \mu_0}{1+\alpha^2}$, $P$ is the polarization factor, $M_s$ is the magnetization saturation, $I$ is the current flowing through the MTJ volume $V$, $m_p$ is the pinned-layer unitary polarization direction and $\beta$, $\epsilon$ and $\epsilon'$ refer to the STT field parameters \cite{Donahue1999a}. The effective magnetic field for a Perpendicular Magnetic Anisotropy (PMA) is defined by the anisotropy field, the external field and the thermal induced field $\boldsymbol{H_{eff}} = \boldsymbol{H_{ani}} + \boldsymbol{H_{ext}} + \boldsymbol{H_{th}}$. The thermal fluctuations induced field is expressed as
\begin{eqnarray}
\boldsymbol{H_{th}} = \boldsymbol{\mathcal{N}(0, 1)} \sqrt{\frac{2 K_B T \alpha}{\gamma' M_s V \Delta_t}}
	\label{eq:hth}
\end{eqnarray}
where $K_B$ is the Boltzmann constant and $\boldsymbol{\mathcal{N}(0, 1)}$ is a Gaussian random vector with components in $\boldsymbol{x, y, z}$ meeting conditions from \cite{Borjas2010, Ament2016}. 

\subsection{Fokker-Plank Equation}
\label{sec:fpe}
Fokker-Planck (advection-diffusion) Equation has been widely used as a tool to establish the probability density function (PDF) of the MTJ magnetization at a given time \cite{Butler2012, Xie2017, Torunbalci2018a, Daniel2021}, being the SDE formalized as 
\begin{eqnarray}
	\frac{\partial \rho}{\partial \tau} &= -\frac{1}{sin(\theta)}\frac{\partial}{\partial \theta} \left[ sin^2(\theta)(i - h -cos(\theta))\rho - \frac{sin(\theta)}{2 \Delta}\frac{\partial \rho}{\partial \theta} \right] \nonumber \\
	 &= \frac{\partial}{\partial \theta} \left[ U(\theta) \rho + D(\theta)\frac{\partial \rho}{\partial \theta} \right]
	\label{eq:fp}
\end{eqnarray}
where $i=\frac{I}{I_c}, h=\frac{\boldsymbol{H_{ext\_z}}}{H^{eff}_k}$ are constant vectors with zero $x, y$ components, $\tau=\frac{t}{\tau_d}$, $I_c=\frac{\alpha H^{eff}_k}{\epsilon \beta}$, $\tau_d=\frac{1}{\alpha \gamma' H^{eff}_k}$ and $H^{eff}_k$ is the effective $z$-component of the shape, interfacial, bulk and voltage-controlled anisotropy. As seen in Equation (\ref{eq:fp}), FPE definition does not involve any stochastic fields. The advection $U(\theta)$ term is responsible for the drift of the distribution while the finite-temperature effects are determined by its initial state $\rho(\theta)|_{\tau=0}=\rho_0(\theta)$, and within its diffusion term $D(\theta)$ \cite{Hu2019a}.

\section{Implemented FPE Solvers}
\label{sec:solvers}
Prior work numerically solves the FPE through finite differences or finite element methods \cite{Butler2012, Xie2017, Torunbalci2018a} or analytical solutions~\cite{Tzoufras2018}. This work implements two different FPE solvers. The first is a numerical FVM approach, which guarantees the conservative properties over the computed magnetization flux \cite{HundsdorferW.andVerwer2007, Daniel2021}. This method solves on non-uniform meshes using an adaptive \emph{upwinding} exponential fitting scheme for the diffusion coefficient, and combines explicit and implicit methods through Crank-Nicolson, preserving stability while increasing accuracy \cite{HundsdorferW.andVerwer2007}.

The FVM solver's computational load limits its practical applicability to thermal regime scenarios where $I<<I_c$. To circumvent this limitation, this work implements a second solver following~\cite{Tzoufras2018}, where an analytical solution to Equation~(\ref{eq:fp}) is presented after expanding $\rho$ as a Legendre polynomial series

\begin{eqnarray}
    \frac{\partial \rho}{\tau} = \sum^{\infty}_{n=0}\sum^{2}_{k=-2} r_n a_{n+k, n} P_{n+k}.
    \label{eq:analytical_0}
\end{eqnarray}

Limiting the series to its first $N$ coefficients  $a_{i, j} j \in [0, N]$,  we can form the pentadiagonal matrix $A$ \cite{Tzoufras2018} leading to:
\begin{eqnarray}
    \frac{\partial \boldsymbol{r(\tau)}}{\partial \tau} = \boldsymbol{A} \boldsymbol{r(\tau)} \implies \boldsymbol{r}(\tau) = e^{\boldsymbol{A} \tau}\boldsymbol{r(0)},
    \label{eq:analytical_1}
\end{eqnarray}
where $\boldsymbol{r}(\tau)$ are the Legendre expansion coefficients of $\rho(\tau)$, and $\boldsymbol{r(0)}$ are the Legendre expansion coefficients of $\rho_0$, the magnetization initial state. The system described by Equations (\ref{eq:analytical_0}, \ref{eq:analytical_1}) is implemented using \scipy~\cite{Virtanen2020}, allowing the computation of long FPE evolutions with simple matrix exponentiation and multiplication operations. 

\begin{figure}[!t]
\centering
\includegraphics[width=\columnwidth]{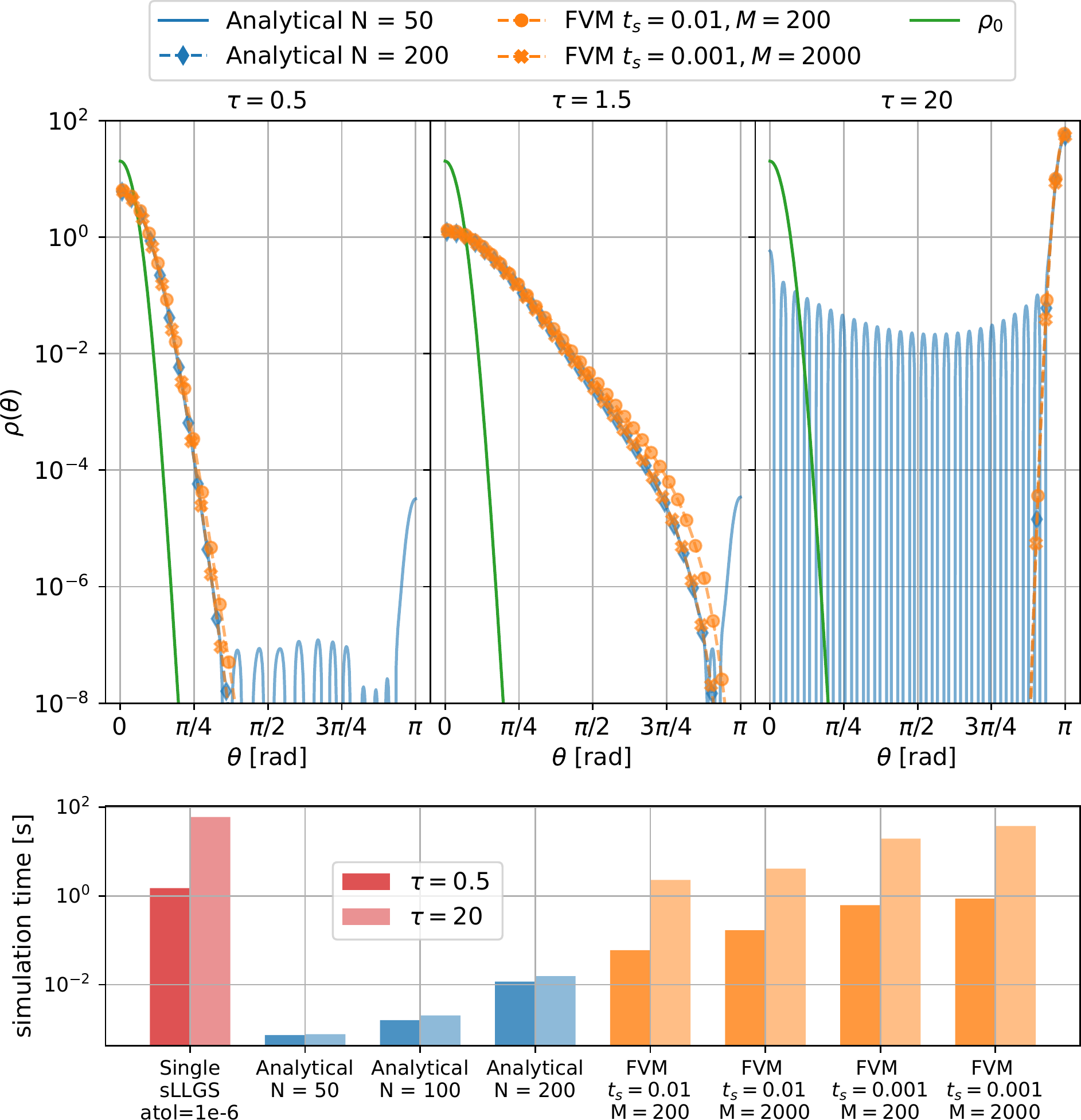}
\caption{
   FPE solvers analysis for an MTJ with $H^{eff}_k=177415\frac{A}{m}$, $\Delta=63$, $\alpha=0.01$,  $M_s=1.2e6\frac{A}{m}$. The top graph shows the effect of resolution on the temporal evolution of Eq. (\ref{eq:fp}). In the bottom graph the FPE computation times are compared against a single \sLLGS transient.
}
\label{fig:comparison}
\end{figure}

Figure~\ref{fig:comparison} shows the evolution of $\rho(\theta)$, switching from $0$ to $\pi$ in response to applied current. With sufficient resolution, both solvers produce closely matched results. Runtime comparison shows that both FPE solvers, even at high resolution, are well below the runtime of a single \sLLGS transient evaluation -- accurate statistical simulation for low error rates would need billions of such evaluations. The implemented solvers therefore offer a practical tool for evaluating MTJ statistics over a wide accuracy/runtime span. The analytical FPE accuracy gets determined by the number $N$ of coefficients, being $N~200$ sufficient, and then computationally-scaling $O(1)$ with the simulated time independently of the $\theta$ dimension. On the contrary, the accuracy of FVM FPE relies on the grid $t_s$ time and $\frac{\pi}{M}$ theta precision -- which involves $O(n^2)$ runtime.

\section{Case Study}
\label{sec:case_study}

\begin{figure}[!t]
\centering
\includegraphics[width=\columnwidth]{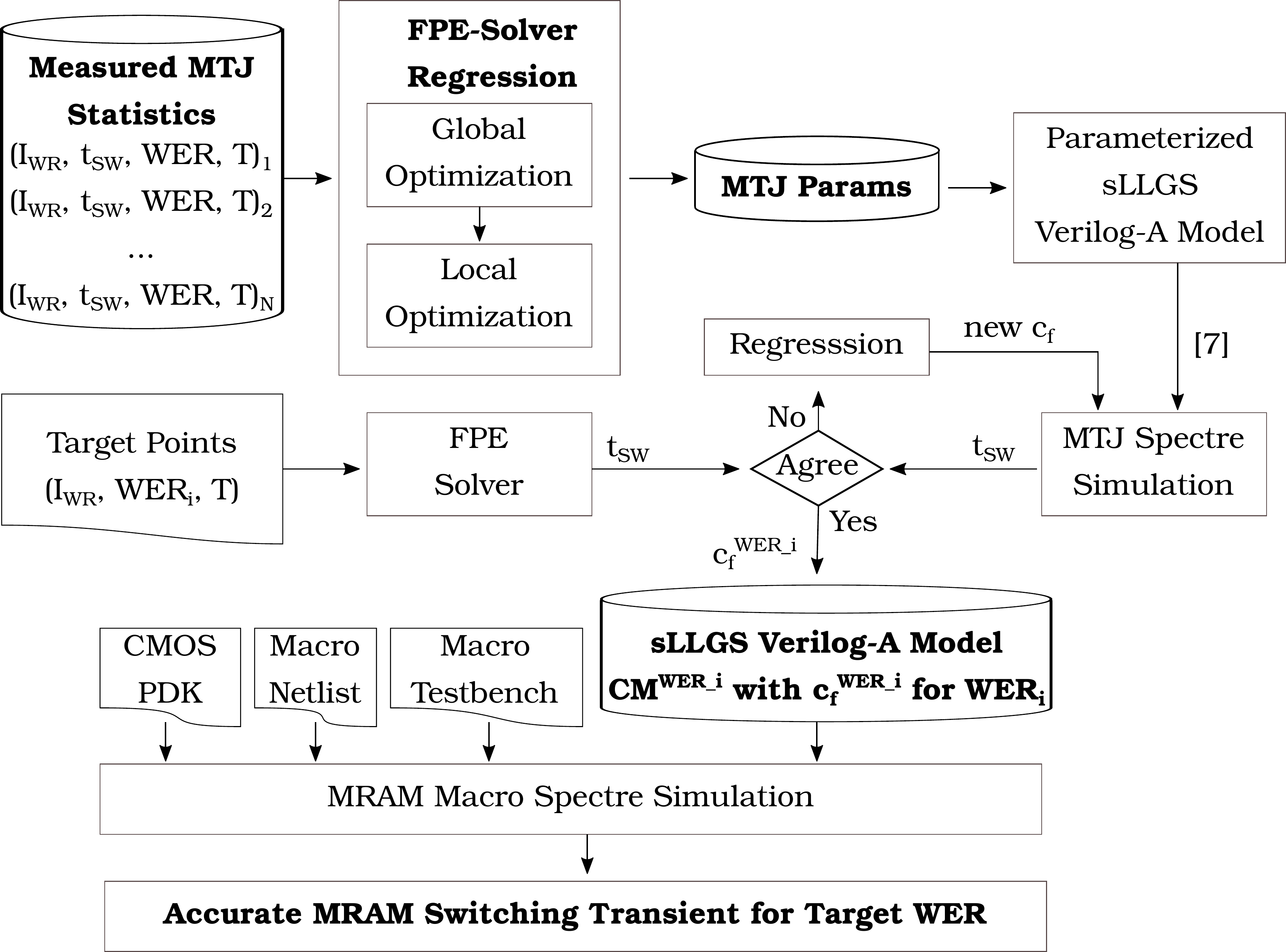}
\caption{
	Proposed methodology for MTJ stochastic behavior analysis and accurate \sLLGS Verilog-A model parameterization for WER switching transient simulation.
}
\label{fig:regression}
\end{figure}

WER/RER dependence on MTJ parameters has been actively investigated, and works like  \cite{Xie2017, Tzoufras2018} emphasize the importance of FPE as its underlying tool. Circuit designers require calibrated MRAM compact models reflecting the behavior of MTJ technologies at different statistical operating points: mean behavior or WER$_{0.5}$, WER$_{1e^{-6}}$ or WER$_{1e^{-8}}$ stochastic corner behaviour. Finding the optimum set of physical parameters that best fit a collection of WER points as a function of a current pulse width and amplitude can be seen as a complex NP-hard problem where a black-box module computes the required time-to-switch under a given current.

Figure~\ref{fig:regression} describes the methodology addressing these problems. For the simpler characterization problem, the FPE solver computes the statistical behavior of a given MTJ operating under known current, pulse width and temperature conditions. On the other hand, the full regression problem requires finding the MTJ parameters that fit known \emph{measured} MTJ statistics -- current/time/temperature switching behaviors for measured WER points. Global optimization algorithms followed by local iterations explore the design space minimizing a target curve.

Closing the loop between FPE and the regression problem, our framework implements an optimization module based on heuristic algorithms using \scipy \texttt{optimize} toolbox~\cite{Virtanen2020} which outputs the best suited MTJ physical parameters. By applying~\cite{Garcia-Redondo2021} the corresponding circuit compact models $\{$CM$_{WER}^i\}$ describing the cell switching transient for WER$_i$ operation points are generated, enabling the circuit designers to simulate such key events. The parameterized models in $\{$CM$_{WER}^i\}$ share the same physical parameters, and only differ in the fitting parameters 
$c_f^{WER\_i}$
thermal fitting coefficients, referring to the fake thermal stress. The expansion of Equation~\ref{eq:sllgs} in spherical coordinates describes $\boldsymbol{m_\theta}$ evolution as proportional to $\boldsymbol{H_{eff \phi}} + \alpha \boldsymbol{H_{eff \theta}}$, leaving $\frac{d}{dt}\boldsymbol{m_\theta} \simeq \frac{\gamma'}{1+\alpha^2} \boldsymbol{H_{eff \phi}}$. The model in \cite{Garcia-Redondo2021} introduces a fictitious $\boldsymbol{H_{fth}}$ term into $\boldsymbol{H_{eff \phi}}$ to emulate the required statistical $\boldsymbol{H_{th}}$ contribution. The emulated $\boldsymbol{H_{fth}}$ is defined as 
\begin{equation}
	\boldsymbol{H_{fth}} = c_f  \sqrt{\frac{2 K_B T \alpha}{\gamma' M_s V \Delta_t}} \boldsymbol{\phi}.
\end{equation}
A straightforward calibration involving a negligible amount of \sLLGS simulations leads to the set of
$c_f^{WER\_i}$
coefficients.

\begin{figure}[!t]
\centering
\includegraphics[width=\columnwidth]{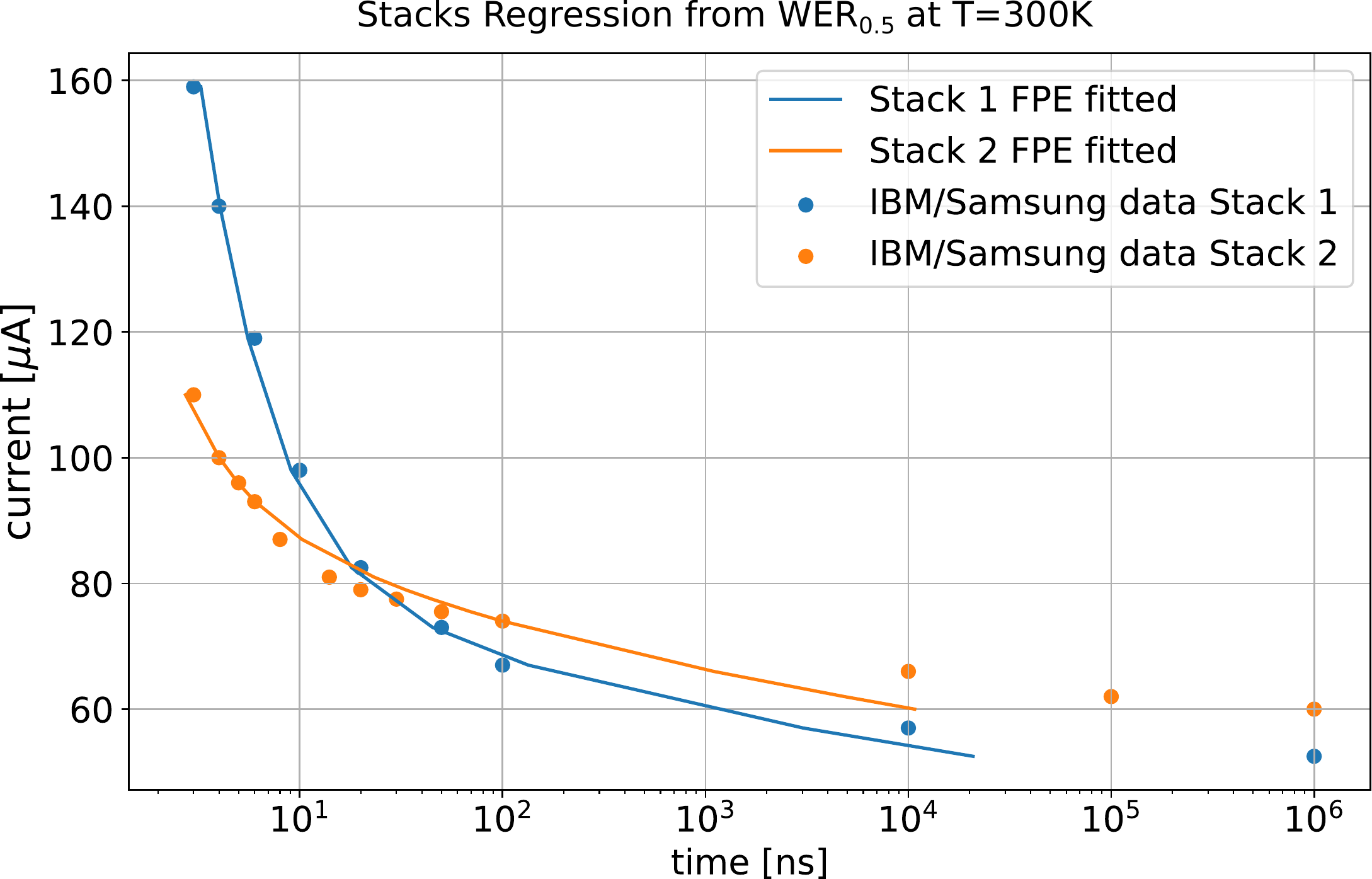}
\caption{
	 Stacks 1 and 2 for fast switching MTJs presented in \cite{Hu2019a}, regressed using the proposed framework.
}
\label{fig:fitting_1}
\end{figure}

\begin{figure}[!t]
\centering
\includegraphics[width=\columnwidth]{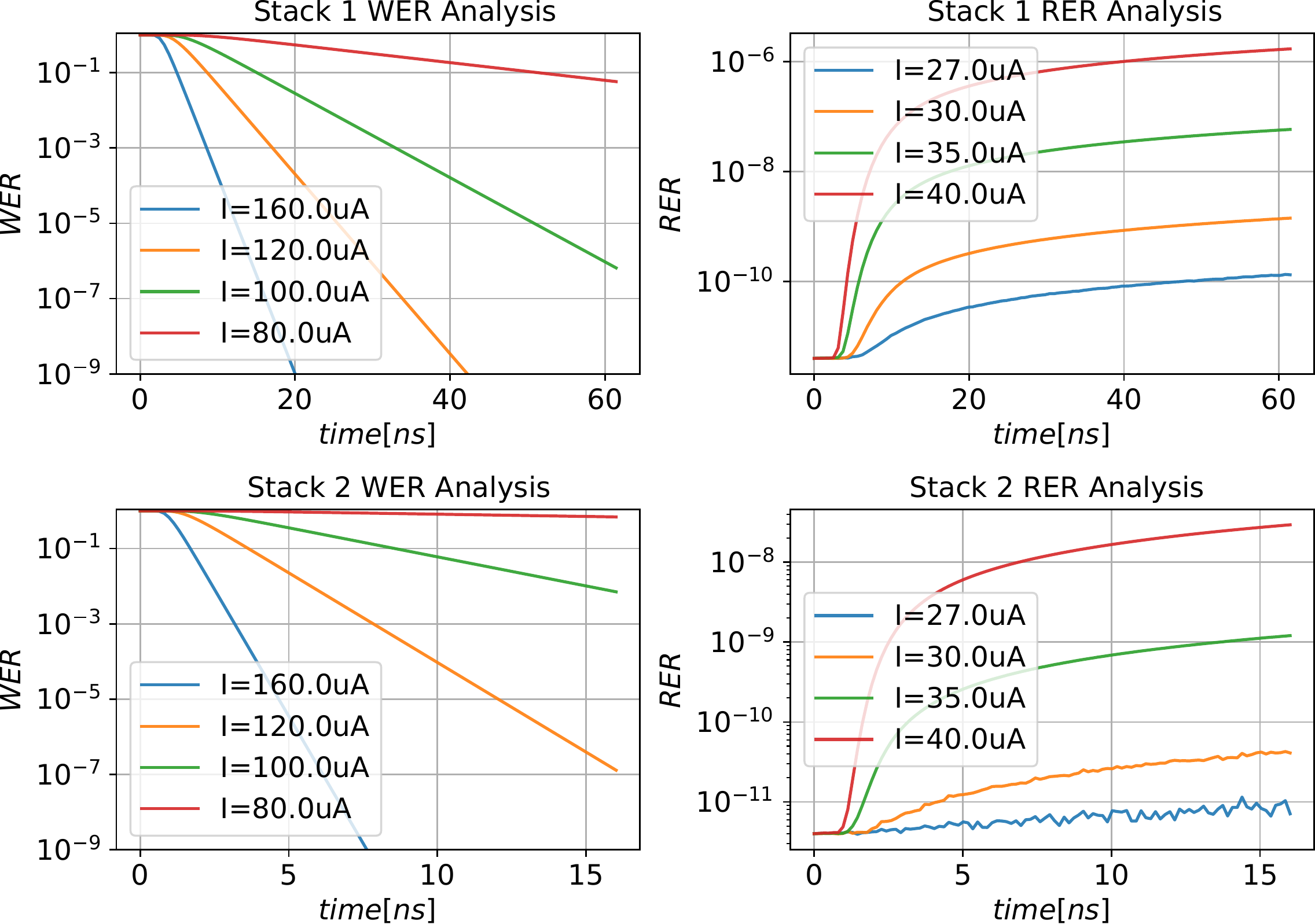}
\caption{
	WER/RER analyses for regressed MTJ stacks \cite{Hu2019a}.
}
\label{fig:wer_rer}
\end{figure}

A case study using foundry data published in \cite{Hu2019a} is presented as an applied example. It is worth noticing that though the included \emph{basin hopping} or \emph{simulated annealing} algorithms attempt to find the global minimum efficiently, as in any heuristic achieving good solutions requires sophisticated problem-tailored cooling/sampling schedules. Figure~\ref{fig:fitting_1} describes the regression results after using the optimizer module -- basin hopping global optimization followed by \emph{L-BFGS-B} algorithm -- fitting to the available WER$_{0.5}$ data points. In this example, the optimization algorithm emphasized the high-current regime as the analyzed stacks were engineered and tested for fast switching~\cite{Hu2019a}.

Once the MTJ parameters are regressed, their statistical behaviour can be studied under any conditions using the FPE solvers as shown in Figure \ref{fig:wer_rer}. WER and RER rates are calculated for different writing and reading currents. It can be noted how Stack 2 is characterized by an easier write-ability, requiring less time to switch in the same current. At low-current regimes Stack 2 tolerates larger currents before flipping the cell during read operations. Such analysis quantifies the impact of MTJ stochasticity taking into account its complex parametric, temperature and current dependencies, helping circuit designers to make informed choices about read/write currents, pulse widths and ECC requirements.

\begin{figure}[!t]
\centering
\includegraphics[width=\columnwidth]{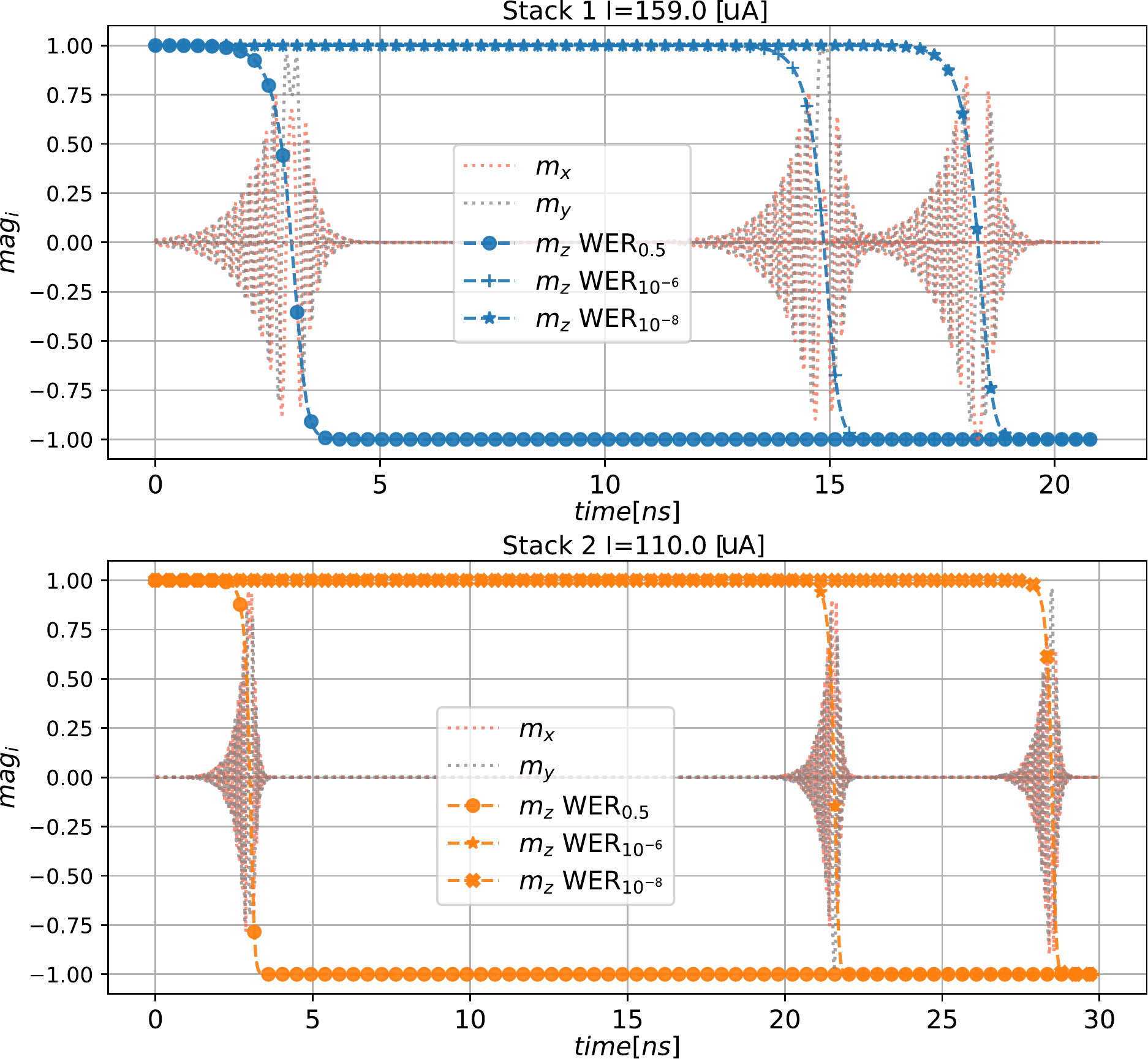}
\caption{
	Transient evaluation for both Stack 1 and 2 \cite{Hu2019a} 
	at WER$_{0.5}$, WER$_{10^{-6}}$ and WER$_{10^{-8}}$ using ~\cite{Garcia-Redondo2021} compact model.
	The figure describes the magnetization vector $\boldsymbol{m}(t)$ decomposed on its three $\boldsymbol{x}, \boldsymbol{y}$ and $\boldsymbol{z}$ components.
}
\vspace{-0.4cm}
\label{fig:transients}
\end{figure}

Finally, making use of the regressed parameters and time/current dependence for WER$_{0.5}$, WER$_{10^{-6}}$ and WER$_{10^{-8}}$, the $c_f^{WER\_i}$ thermal fitting coefficients are calibrated using single MTJ \sLLGS transient simulation to generate circuit-ready transient models. Figure~\ref{fig:transients} depicts the simulated transients of Stack 1 and Stack 2 cells accurately describing the switching events.

\section{Conclusions}
\label{sec:conclusions}
This work presented a framework for the characterization and analysis of MTJ stochasticity. We implemented and analyzed two FPE solvers (numerical FVM and analytical), and presented an optimization module that orchestrates the efficient computation of MTJ statistics and parameter regression. Finally, we applied the proposal in a case study with published foundry data, demonstrating efficient MTJ parameter regression and the generation of \sLLGS Verilog-A models enabling the simulation of target WER switching transients. The framework code is available upon request.

\section{Acknowledgments}
The authors would like to thank Milos Milosavljevic and Cyrille Dray for their helpful discussions.

\vspace{-0.1cm}
\bibliography{essderc_2021}
\bibliographystyle{noUrlIEEEtran}

\end{document}